\documentstyle[aps,preprint]{revtex}
\begin{document}
\draft
\title{
Calculation of Electric Unit charge
}
\author{Il-Tong Cheon}
\address{Department of Physics, Yonsei University, Seoul 120-749,Korea}

\maketitle

\vskip1cm
\begin{abstract}
Considering the stresses due to the vacuum fluctuation and the electric charge loaded
over the surface of a spherical cavity, we estimate the maximum value of the charge.
Since this value is independent of the cavity size and parameter free, it is regarded
as the electric unit charge. Our result is $Q= 1.55\times 10^{-19}$ Coulomb
which implies the relevant fine structure constant $\alpha=1/145.90$.
\end{abstract}


\vspace{10pt}
The most fundamental constants in physics are the speed of light c,
the Planck constant h and the electric unit charge e. Their numerical values,
c=2.997924562(11) $\times 10^{8}$ m/sec, h=6.6260755(40) $\times  10^{-34}$ J.sec
and e=1.60217733(49)$\times 10^{-19}$ Coulomb,
were determined experimentally.
However, physicists should sometime
explain where those values could come from.
If they  could be derived on the purely theoretical basis, it would be
nice and help us to have deeper understanding of nature.

Let us consider a cavity, and suppose that some amount of electric charge is
loaded over its surface. Then, the charge  exerts an outward stress so that the cavity will explode.
on the other hand, it is well known that the vacuum  fluctuation in the cavity yields
 the Casimir force~\cite{ref1}. If this force is inward, the cavity may stay in
 equilibrium on the balance between these two forces. Previous calculation~\cite{ref2,ref3}
 derived an outward force due to vacuum fluctuation for a spherical cavity.
 However, for the case of two parallel plates, the Casimir force is definitely
 inward~\cite{ref1,ref4}.

Although the Casimir force is dependent of the cavity shape, it is hard to
believe such a drastic change in its sign. There might be a pitfall in
the calculation of the Casimir force for a spherical cavity. Notice
that the Casimir force for a cube turns out to be inward in calculation
based on the Casimir's semi-classical treatment~\cite{ref1,ref4}.

In this situation, it is important to estimate again very carefully the
Casimir force for a spherical cavity by a rigorous method extended in this
letter.
If the Casimir force could turn out to be inward, one would be allowed to ask how
 much charge should be loaded on the cavity surface to save it from collapse.
 This amount of charge might be related to the electric unit charge.

 First of all, we shall investigate precautiously a spherical cavity in
 contex of works on various phenomena occuring between two parellel plates[1,4-21].

The zero-point energy in a free space is given by
\begin{equation}
E = 2 \left({\hbar c \over 2}\right) \left ({ a \over {2 \pi} }\right)^{3} \int {\sqrt {k_1^{2} + k_2^{2} + k_{3}^2}}
~dk_x dk_y dk_z
\end{equation}
where $a$ is the size of normalization box. For finite space, the wave number
becomes discrete, i.e. $k_x= \frac{\pi}{a}  n_1 ,~ k_y=\frac{ \pi}{a} n_2 ~$ and
$k_z=\frac{ \pi}{a}   n_3 ~$  where $n_i$ are integers.
Then, the Casimir energy in a spherical cavity of diameter $a$ or a cubic cavity of
$ a \times a \times a $ can be deduced from the zero-point energy of electromagnetic vacuum
field expressed as~\cite{ref1}

\begin{eqnarray}
   E&=&2\left(\frac{\hbar c}{2}\right)\frac{1}{8}
   \sum_{n_1,n_2,n_3 \atop{=-\infty}}^{\infty}\left[\left(\frac{\pi}{a} n_1\right)^2
   +\left (\frac{\pi}{a} n_2\right)^2+\left (\frac{\pi}{a} n_3\right)^2\right]^{\frac{1}{2}}   \nonumber \\
    &=&\frac{\hbar c}{8}\left(\frac{\pi}{a}\right)\sum_{n_1,n_2,n_3
    \atop{=-\infty}}^{\infty}
    [n_1^2+n_2^2+n_3^2]^{\frac{1}{2}}.
\end{eqnarray}

This expression is obtained simply by introducing the discrete photon momentum
which implies the discrete photon propagator. Although this procedure does not
give a complete answer, it provides a good approximation. For the case of finite
space restricted by two parallel plates, Bordag et al.~\cite{ref22} derived the photon
propagator which satisfies the boundary condition on the plate surfaces.
And it was shown~\cite{ref13} that the discrete photon propagator corresponds to the
lowest order of the field theoretical propagator proposed by Bordag et al.~\cite{ref22}.
Therefore, eq.(2) may be regarded as the lowest order of the zero-point energy derived
based on the field theory. The summation is actually divergent, but if the
value in the free space is substracted from it, the rest value will remain finite
and this quantity gives the Casimir energy.

To evaluate the summation, we apply the Poisson's summation formula on Fourier
Transform \cite{ref23},
\begin{equation}
\sum_{n = - \infty}^{\infty} f(n)
= \sum_{s = - \infty}^{\infty} \int_{- \infty}^{\infty} f(x) e^{i \xi_s x} dx,
\end{equation}
with $\xi_s = 2 \pi s$.  This formula suggests that the sum of $f(n)$   can be converted into the
sum of its Fourier transformed function. Validity of eq.(3) can be seen
with $f(n)=\exp(-\pi n^2) $  for which the integral
on the right-hand side yields $\exp(-\pi s^2)$   .
Another example is $f(n)={1 \over (\beta^2+n^2)} $  with $\beta > 0 $  . The integral in eq.(3) is easily evaluated
and, then, we have

\begin{equation}
\sum_{n = - \infty}^{ \infty} { 1 \over { \beta^2 + n^2} }=
\sum_{s = - \infty}^{ \infty} \left ( {\pi \over \beta}\right) e^{ - 2 \pi \beta |s|}.
\end{equation}
Both sides give identical answer,
$ \left ({\pi \over \beta}\right) ( e^{\pi \beta}+e^{-\pi \beta}) / ( e^{\pi \beta}-e^{-\pi \beta})
$ \cite{ref24}.
For a three component function
, the formula (3) is expressed as

\begin{equation}
\sum_{n_1,n_2,n_3 \atop{= - \infty}}^{\infty} f(n_1,n_2,n_3)
= \sum_{\lambda_1,\lambda_2,\lambda_3 \atop{=-\infty}}^{\infty}
 \int_{-\infty}^{\infty} f(p_1,p_2,p_3) \exp[i(\xi_1 p_1+\xi_2 p_2+\xi_3 p_3)]
 dp_1 dp_2 dp_3,
\end{equation}
with $\xi_i=2 \pi \lambda_i$   . Thus, the summation in eq.(2) can be expressed as

\begin{equation}
\sum_{n_1,n_2,n_3\atop{=-\infty} } ^{\infty} \sqrt{ n_1^2+n_2^2+n_3^2}
= \sum_{\lambda_1,\lambda_2,\lambda_3\atop{=-\infty}}^{\infty}
 \int \sqrt{p_1^2+p_2^2+p_3^2} \exp[i(\xi_1 p_1+\xi_2 p_2+\xi_3 p_3)]
 dp_1 dp_2 dp_3.
\end{equation}

In the spherical coordinates, $ n_1=n \sin \theta \cos \phi,~n_2=n \sin \theta \sin \phi,~
n_3=n \cos \theta,$
similarly
$ p_1=p \sin \theta \cos \phi,~p_2=p \sin \theta \sin \phi,
~p_3=p \cos \theta$ and $ \xi_1=\xi \sin \theta^{\prime}  \cos \phi^{\prime} ,
\xi_2=\xi \sin \theta^{\prime}  \sin \phi^{\prime} ,
\xi_3=\xi \cos \theta^{\prime}.$
Then we have
\begin{eqnarray}
\sum_{i} \xi_i p_i &=& p \xi (\sin \theta \cos \phi
\sin \theta^{\prime} \cos \phi^{\prime}+
\sin \theta \sin \phi  \sin \theta^\prime \sin \phi^\prime
+\cos \theta \cos \theta^\prime)  \nonumber \\
&=& p \xi [\sin \theta \sin \theta^\prime \cos(\phi-\phi^\prime)+\cos \theta \cos\theta^\prime]
\nonumber \\
&=& p \xi \cos \omega,
\end{eqnarray}
where $\omega$  is the angle between  $\mbox{\boldmath $p$}$ and $\mbox{\boldmath $\xi$}$.  Accordingly,
eq.(6) becomes
\renewcommand{\theequation}{8.\alph{equation}}
\setcounter{equation}{0}
\begin{eqnarray}
\mbox{left-hand side} &=& \sum_{n=-\infty}^{\infty}
\sqrt{(n \sin \theta \cos\phi)^2+(n \sin \theta \sin \phi)^2+(n \cos \theta)^2}
=\sum_{n=-\infty}^{\infty} \sqrt{\mbox{\boldmath $n$} \cdot
\mbox{\boldmath $n$}},
\\
\mbox{right-hand side} &=& \sum_{\lambda=-\infty}^{\infty} \int
\sqrt{\mbox{\boldmath $p$} \cdot \mbox{\boldmath $p$}} e^{i \xi p \cos \omega}
p^2 dp \sin \theta d\theta d\phi,
\end{eqnarray}
with $\xi=2 \pi \lambda$.
The integral in eq.(8b) is indefinite, i.e.
ultraviolet divergence. However, for wavelengths shorter than the atomic
size, it is unrealistic to use a model of cavity. Therefore, we take
the well known regularization procedure such as
introducing a smooth cut-off function, $e^{-\varepsilon p}$, and making a limit
 $\varepsilon\rightarrow0$.
\setcounter{equation}{8}
\renewcommand{\theequation}{\arabic{equation}}

As a result, the zero-point energy is found in the spherical coordinates as

\begin{equation}
E={{\hbar c} \over 8} \left ({\pi \over a}\right) \sum_{\lambda=-\infty}^{\infty} \lim_{\varepsilon \rightarrow 0}
\int_{0}^{\infty} dp |\mbox{\boldmath $p$}| p^2 e^{-\varepsilon p} \int_{0}^{2\pi} d\phi \int_{0}^{\pi} d\theta \sin \theta
 e^{i \xi p \cos \omega}.
\end{equation}
For $\lambda=0$  ,  we have

\begin{equation}
E_0={{\hbar c} \over 8} \left({\pi \over a}\right) \lim_{\varepsilon \rightarrow 0}
\int |\mbox{\boldmath $p$}| e^{-\varepsilon p} p^2 dp \sin \theta d\theta d\phi
=\hbar c \left ({a \over 2 \pi}\right)^3 \int |\mbox{\boldmath $k$}|d\mbox{\boldmath $k$},
\end{equation}
where  $k_i=\left ({\pi \over a} \right) p_i$ is used. This value is exactly identical to the zero-point energy in a free
space with a normalization volume $a^3$ .  The Casimir energy is,
then, obtained for $\lambda \neq 0$
by expanding the exponential in eq.(9) with the spherical hamonics as

\begin{eqnarray}
{\cal E}_c &=&{{ \hbar c} \over 8} \left ({ \pi \over a}\right)
\sum_{\lambda=-\infty,\atop{(\lambda \neq 0)}}^{\infty} \lim_{\varepsilon \rightarrow 0}
4 \pi \sum_{l,m} i^l \int _{0}^{\infty}p^3 e^{-\varepsilon p} j_l (\xi p) dp
Y_{lm}^{*} (\theta^\prime,\phi^\prime)\int Y_{lm}(\theta,\phi) \sin\theta d\theta d\phi
\nonumber \\
&=&{{\hbar c}\over{8}}\left ( {{\pi}\over{a}} \right ) \sum_{\lambda=-\infty, \atop{(\lambda\neq 0)}}^{\infty}
\lim_{\varepsilon\rightarrow 0} 4 \pi \int_{0}^{\infty} p^3 e^{-\varepsilon p}
j_0(\xi p) dp \nonumber \\
&=& {{\hbar c} \over 8} \left ({\pi \over a}\right) \sum_{\lambda=-\infty,\atop{(\lambda \neq 0)}}^{\infty}
\lim_{\varepsilon \rightarrow 0} \frac{8 \pi (3 \varepsilon^2-\xi^2)}{(\varepsilon^2+\xi^2)^3}
\nonumber \\
&=& -{{\hbar c} \over 8} \left (\frac{\pi}{a}\right) {1 \over \pi^3} \sum_{\lambda=1}^{\infty} {1 \over \lambda^4},
\end{eqnarray}
where the summation is just the Riemann's zeta function,$\zeta(4)=\pi^4/90$     .
Thus, for $\lambda \neq 0 $ , we find
\begin{equation}
{\cal E}_c=-\frac{\hbar c}{8 \pi^2 a} \zeta(4).
\end{equation}
This is the Casimir energy of a spherical cavity with diameter $a$. Notice the
negative sign!

At this stage, we explore why the previous calculation yielded the Casimir
energy with a positive sign.
After rewriting the integral in eq.(1) as
\begin{eqnarray}
\int |\mbox{\boldmath $k$}| d\mbox{\boldmath $k$}&=&4\pi \int_{0}^{\infty} |\mbox{\boldmath $k$}| k^2 dk
= 2\pi\int_{-\infty}^{\infty} |\mbox{\boldmath $k$}| k^2 dk \nonumber \\
&\rightarrow& 2\pi \left (\frac{\pi}{a}\right)^4 \sum_{n=-\infty}^{\infty} |\mbox{\boldmath $n$}| n^2,
\end{eqnarray}
we apply the formular, (3), to obtain
\begin{equation}
2\pi \left ({\pi \over a}\right)^4 \sum_{n=-\infty}^{\infty} |\mbox{\boldmath $n$}| n^2=2\pi\left ({\pi\over a}\right)^4
\sum_{s=-\infty}^{\infty} \int_{-\infty}^{\infty}|\mbox{\boldmath $p$}| p^2 e^{i \xi_s p} dp,
\end{equation}
where $\xi_s=2\pi s$  . For $s=0$  , we find the right-hand side as
\begin{eqnarray}
2\pi\left ({\pi\over a}\right)^4 \int_{-\infty}^{\infty} |\mbox{\boldmath $p$}| p^2 dp
&=& \left ({\pi\over a}\right)^4 4\pi \int_{0}^{\infty}|\mbox{\boldmath $p$}| p^2 dp \nonumber \\
&=&\left ({\pi\over a}\right)^4 \int |\mbox{\boldmath $p$}| d\mbox{\boldmath $p$}=\int |\mbox{\boldmath $k$}|d\mbox{\boldmath $k$}.
\end{eqnarray}
This is exactly identical to the quantity in a free space.
For $s\neq0$  , eq.(14) is evaluated as
\begin{eqnarray}
2\pi \left ({\pi\over a}\right)^4 \sum_{s=-\infty,\atop{(s \neq 0)}}^{\infty}
\int_{-\infty}^{\infty} |\mbox{\boldmath $p$}|p^2 e^{i \xi_s p} dp
&=& 4\pi \left ({\pi \over a}\right)^4 \sum_{s=1}^{\infty} \int_{-\infty}^{\infty}
|\mbox{\boldmath $p$}| p^2\cos(\xi_s p) dp \nonumber \\
&=& 8\pi \left ({\pi \over a}\right)^4 \sum_{s=1}^{\infty}\lim_{\varepsilon\rightarrow0}
\int_{0}^{\infty} p^3 e^{-\varepsilon p} \cos(\xi_s p) dp \nonumber \\
&=& 8 \pi \left ({\pi \over a}\right)^4 \sum_{s=1}^{\infty} \lim_{\varepsilon\rightarrow0}
\frac{6(\varepsilon^4-6\varepsilon^2\xi_s^2+\xi_s^4)}{(\varepsilon^2+\xi_s^2)^4}
\nonumber \\
&=& 48\pi \left ({\pi\over a}\right)^4 \sum_{s=1}^{\infty}{1\over \xi^4} \nonumber \\
&=& \left ({\pi \over a}\right)^4 {3\over \pi^3} \zeta(4).
\end{eqnarray}
Thus, we find
\begin{equation}
{\cal E}_c=\frac{\hbar c}{8} \left (\frac{\pi}{a}\right) {3 \over \pi^3} \zeta(4)=
\frac{3 \hbar c}{8 \pi^2 a} \zeta(4)=\frac{\pi^2 \hbar c}{240 a}.
\end{equation}
This result has, indeed, a positive sign and is consistent with other previous
 calculations~\cite{ref2,ref3}. However, it is controversial because of
 one-dimensional calculation
 in principle. It should be calculated three-dimensionally as was presented through
 eqs.(6)-(12). The factor, ~$\exp(i \xi p \cos\omega)$,    in eq.(9) is actually a key point to obtain a negative
 sign.
 Although one dimensional calculation is run with the factor,
 $\exp(i \xi_s p )$
 which is independent of angles, three dimensional calculation
 contains the angle $\omega$
 in the factor which is effective to the integrals over $\theta$   and
 $\phi$  . However,
 after integration over angles  $\theta$  and $\phi$  ,
 the result does not have any dependence of
 angles $\theta^\prime$  and $\phi^\prime$ at all, because of
 $Y_{00}(\theta^\prime,\phi^\prime)={1 \over \sqrt{4 \pi}}$. As a result,
 $\cos(\xi_s p)$ in the integral over p appearing in eq.(16)
  is replaced by  $j_0(\xi p)$ as seen in eq.(11). Thus, the negative sign is realized.

Finally, the stress due to the vacuum fluctuation is obtained by
differentiating eq.(12) with respect to $a$ as
\begin{equation}
{\cal P}_c=\frac{1}{4\pi(a/2)^2}\left [-\frac{\partial {\cal E}_c}{\partial a}\right ]
=-\frac{{\hbar} c }{8 \pi^3 a^4} \zeta(4)=-\frac{\pi {\hbar} c}{720 a^4}.
\end{equation}
If there were only this inward force acting on the cavity surface, it would
collapse.

How much electric charge do we have to load over the cavity surface to stabilize it?

Supposing that the electric charge of $Q$ is loaded and applying the Gauss's law to the sphere,
we obtain the electric field in the normal direction to the surface,
\begin{equation}
E_n=\frac{4\pi Q}{4\pi(a/2)^2}=\frac{4 Q}{a^2}.
\end{equation}
Then, the stress due to charge Q can be found as\cite{ref25}

\begin{equation}
{\cal P}_e=\frac{E_n^2}{8\pi}=\frac{2 Q^2}{\pi a^4}.
\end{equation}
From the stability condition, ${\cal P}_c+{\cal P}_e=0$     , we obtain
\begin{equation}
Q=\left (\frac{\pi^2 \hbar c}{1440}\right)^{1\over2}=1.55\times 10^{-19}~ Coulomb.
\end{equation}
It is surprising that  the electric charge is completely independent
of the cavity size. This result might be retained even in the limit,
$a\rightarrow0$    .
Therefore, it may be regarded as the electric unit charge.

Accordingly, the value associated with the fine structure constant is

\begin{equation}
\alpha={1 \over 145.90}.
\end{equation}
The experimental value of the fine structure constant is~\cite{ref26}

\begin{equation}
\alpha_{exp}={1 \over 137.035987(29)}.
\end{equation}
Discrepancy between $\alpha$  and $\alpha_{exp}$  is only 6 \%  . This discrepancy is not much important here.
The essential  point is that the $\alpha$  value has been firstly calculated based on purely theoritical analysis.

If we start with the fully field theoretical photon spectrum, the result may be
improved. It will be explored on another occasion.

\acknowledgements

This work was supported by the Korean Ministry of Education(Project no.
BSRI-95-2425). The author thanks Dr. Su Houng Lee for valuable discussion.

\end{document}